\documentclass[a4paper,12pt]{article}
\usepackage{amsmath}
\usepackage{amsfonts}
\usepackage{amssymb}
\usepackage{color}
\usepackage{graphicx} 
\usepackage{graphics} 

\title{The fate of Lorentz frame\\ in the vicinity of black hole singularity}
\author{
Douglas G. Moore{\footnote{Douglas{\textunderscore}Moore1@baylor.edu}}\,\, and 
V. H. Satheeshkumar{\footnote{V{\textunderscore}HSatheeshkumar@baylor.edu}}\\ 
Department of Physics, Baylor University,\\ Waco, TX, 76798-7316, USA. 
}
\date{}
 \newcommand{\bq}{\begin{equation}}
 \newcommand{\eq}{\end{equation}}
 \newcommand{\bqn}{\begin{eqnarray}}
 \newcommand{\eqn}{\end{eqnarray}}

\begin{document}
\maketitle

\begin{abstract}

General Relativity is known to break down at singularities. However,
it is expected that quantum corrections
become important when the curvature is of the order of Planck scale
avoiding the singularity. By calculating the effect of tidal forces on a
freely falling inertial frame, and assuming the least possible size of
the frame to be of the Planck length, we show that the Lorentz frames
cease to exist at a finite distance from the singularity. Within that
characteristic radius, one cannot apply General Relativity nor Quantum
Field Theory as we know them today. Additionally we consider other quantum
length scales and impose limits on the distances from the singularity at
which those theories can conceivably be applied within a Lorentz frame.
\end{abstract}

\vfill

\begin{center}

\textbf{This essay received an \textit{honorable mention} in the Gravity Research Foundation 2013 Awards for Essays on Gravitation.}
\end{center}

\vfill

\newpage
\section{Introduction}

Lorentz invariance is built into the traditional Quantum Field Theories (QFT) by construction, whereas General Relativity (GR) has Lorentz invariance as a local symmetry through the Equivalence Principle. The \textit{strong} Equivalence Principle (SEP) states that in a freely falling, non-rotating laboratory occupying a small region of spacetime, the laws of physics are those of Special Relativity (SR). Further, as it is in Newtonian gravity, the acceleration of such a freely falling object (which we call an \textit{Einstein elevator} or simply an \textit{elevator}) can be constant only over short distances. While considering the large distances in a gravitational field, one has to take the tidal forces into account. While these tidal forces are ever present in a finite size elevator, they may be negligible given the scale of the physics in consideration. We use this fact to determine the lower limit on the mass of a spherical symmetric black hole\footnote{These considerations apply equally well to any other type of black hole; though due to the significant increase in complexity, we focus only on the Schwarzschild case herein. Additional consideration of other types will be provided in \cite{MS}} in order for the tidal forces on a fiducial elevator of Planck size at the event horizon to be negligible. We assert that mass as the lower bound on the masses that GR can handle; for any mass lower than this, the fiducial elevators will break down before reaching the horizon and would then be observable. As we apply the same analysis to other length scales, for example, Quantum Electrodynamics (QED), we find higher lower-bound masses and radii at which the Einstein elevator must begin taking tidal corrections. This corresponds to the breakdown of Rindler-space at the horizon and thus has an immediate affect on some derivations in black hole thermodynamics \cite{Susskind}. Further still, since the usual Quantum Field Theories have Minkowski spacetime as their background, we see these distances as the limit beyond which the traditional QFT in question fails and must either begin considering curved spacetimes as their background or be altered in some other fundamental way.

\section{Equation of Geodesic Deviation }
\renewcommand{\theequation}{2.\arabic{equation}} \setcounter{equation}{0}

In GR, the tidal force on an object by a gravitating body is given by the equation of geodesic deviation.\footnote{For a detailed discussion please see the reference \cite{HEL}. } 
For an observer in an orthonormal freely falling frame, the geodesic deviation equation is given by
\begin{equation}
\frac{d^2 \zeta^{\hat{\alpha}} }{d \lambda^2} = c^2 R^{\hat{\alpha}}_{\,\,\,\,\, \hat{0} \hat{0} \hat{\delta} } \zeta^{\hat{\delta}}
\end{equation}
where $\zeta^{\hat{\alpha}}$ is the orthonormal connecting vector joining two points on the neighboring geodesics parameterized by some affine parameter $\lambda$ and $R^{\hat{\alpha}}_{\,\,\,\,\, \hat{\beta} \hat{\gamma} \hat{\delta} }$ are components of the Riemann tensor in the tetrad frame which can be expressed in terms of the Riemann tensor components in the coordinate basis and the tetrads as
\begin{equation}
R^{\hat{\alpha}}_{\,\,\,\,\, \hat{\beta} \hat{\gamma} \hat{\delta} } = R^{\mu}_{\,\,\,\,\, \sigma \nu \rho  } \left( \hat{e}^\alpha \right)_\mu  \left( \hat{e}_\beta \right)^\sigma \left( \hat{e}_\gamma \right)^\nu \left( \hat{e}_\delta \right)^\rho. 
\end{equation}

\section{Tidal Forces of a Schwarzschild Black Hole}
\renewcommand{\theequation}{3.\arabic{equation}} \setcounter{equation}{0}

The complete gravitational collapse of a body always produces a Kerr-Newman black hole \cite{Hawking-Ellis}. Such a black hole is characterized by only three parameters namely its mass, electric charge and angular momentum. By appropriately taking one or two of these parameters to be zero, we can obtain all the four\footnote{Schwarzschild (1916), Reissner-Nordstr\"{o}m (1918), Kerr (1963) and Kerr-Newman (1965)} possible asymptotically flat Black Hole solutions in GR \cite{Chandra}. Since tidal force expressions are more complicated for a Kerr-Newman case they are considered elsewhere \cite{MS}. In this essay we limit our discussions to only the Schwarzschild black hole.

The Schwarzschild metric in the usual coordinates is given by
\bq
ds^2= \left( 1-\frac{2\mu}{r}\right) c^2 dt^2 - \left( 1-\frac{2\mu}{r}\right)^{-1} dr^2 - r^2 (d\theta^2 + sin\theta^2 d\phi^2),
\eq
where $\mu = \frac{GM}{c^2}$.

The standard vierbein $e_m^{\,\,\,\, \mu}$ for Schwarzschild metric is defined as
\[
e_m^{\,\,\,\, \mu} =
 \begin{pmatrix}
  \left( 1 - \frac{2 \mu}{r} \right)^{-1/2} & 0 & 0 &  0 \\
  0 & \left( 1 - \frac{2 \mu}{r} \right)^{1/2} & 0 & 0 \\
  0 & 0 & \frac{1}{r} &  0 \\
    0 & 0 & 0 & \frac{1}{r\, \sin\theta}
 \end{pmatrix}
 \]
and its inverse vierbein $e^m_{\,\,\,\, \mu}$ is
 \[
e^m_{\,\,\,\, \mu} =
 \begin{pmatrix}
  \left( 1 - \frac{2 \mu}{r} \right)^{1/2} & 0 & 0 &  0 \\
  0 & \left( 1 - \frac{2 \mu}{r} \right)^{-1/2} & 0 & 0 \\
  0 & 0 & r &  0 \\
    0 & 0 & 0 & r\, \sin\theta
 \end{pmatrix}.
 \]

Using the above, the computed components of the tidal force are given by

\bqn
\frac{d^2 \zeta^{\hat{r}} }{d \lambda^2}&=& + \frac{2\mu}{r^3} c^2 \zeta^{\hat{r}} \\ 
\frac{d^2 \zeta^{\hat{\theta}} }{d \lambda^2} &=& - \frac{\mu}{r^3}  c^2 \zeta^{\hat{\theta}} \\
\frac{d^2 \zeta^{\hat{\phi}} }{d \lambda^2} &=& - \frac{\mu}{r^3} c^2 \zeta^{\hat{\phi}} 
\eqn

It is clear from the signs of equations that the tidal forces stretch the infalling object in the radial direction and squeeze in the angular directions. 

\section{Theoretical Limits}
\renewcommand{\theequation}{4.\arabic{equation}} \setcounter{equation}{0}

From the dimensions of the tidal force equations, we know that these equations give us the acceleration gradient experienced by the freely falling observer. In other words, they have the dimensions of inverse time squared. We can have the dimensions of inverse length squared by dividing both the sides of the equations by $c^2$. Considering only the radial component and taking $L$ be the length scale at which the tidal forces of the black hole kick in, we have after substituting for $\mu$,

\bq
\frac{1 }{L^2} =  \frac{2 G M}{c^2 r^3}.
\eq

 If $l$ is the length scale of the experiment, then for tidal forces not to have an effect on the experiment, we should have $ l < < L.$ That is,
 
\bq
 \frac{1 }{l^2} >> \frac{2 G M}{c^2 r^3}.
 \eq

This implies that the freely frame of size $l$ will not feel the tidal forces or stays inertial only if it is at distance $r$ from the singularity which is,
\begin{equation}
\boxed{ r_{min}^3 >> 
r_s l^2.}
\end{equation}
where $r_S = \frac{2 G M}{c^2}$ is the Schwarzschild radius.

These lower limits on distances for a solar mass black hole for various length scales are summarized in the table below.

Since the events inside the event horizon does not affect the physics outside the black holes, we now concentrate on what happens at the Schwarzschild radius, $r_S$, and derive limits on the mass of the black hole. At the horizon,

\bq
 \frac{1 }{l^2} >> \frac{2 G M}{c^2 r_{S}^3}. 
 \eq

Substituting for $r_S = \frac{2 G M}{c^2}$, we have

\bq
 {l} << \frac{2 G M}{c^2} 
 \eq
  or 
\begin{equation}
\boxed{ M_{min} >> \frac{c^2}{2 G} l .}
\end{equation}

The lower limits on the mass for various length scales are summarized in the table below.

\begin{center}
  \begin{tabular}{|c|c|ccc|ccc|}
    \hline
    Type of    & $ l \sim $         & &$M_{min}\, >>  $&       &  & $r_{min}\, >>$ & \\
Experiment & $m$  & $kg  $ & $  M_{Sun} $ & $ m_p $    & $  m  $& $r_S$ & $  l_p  $   \\\hline   
     QED        & $  10^{-12}$ & $ 10^{14}$ & $ 10^{-16}$ & $10^{22}$ & $ 10^{-21}$ & $ 10^{-24}$ & $10^{14}$ \\\hline   
    QCD        & $ 10^{-15}$ & $10^{11}$ & $ 10^{-19}$ & $10^{19}$ & $10^{-27}$ & $10^{-30}$ & $10^{8}$ \\\hline   
    Electroweak         & $ 10^{-18}$ & $ 10^{8} $ & $ 10^{-22}$ & $10^{16}$ & $ 10^{-33}$ & $10^{-36}$ & $10^{2}$ \\\hline   
    Planck     & $ 10^{-35}$ & $ 10^{-9}$ & $ 10^{-39}$ & $10^{-1}$ & $ 10^{-67}$ & $10^{-70}$ & $10^{-32}$ \\\hline   
  \end{tabular}
\end{center}

Clearly these masses are significantly less than a solar mass, but should become significant in scenarios involving evaporation. In effect, one cannot consider the space in the vicinity of the horizon to be Rindler giving us a measure of what it means for a black hole to be ``big".

Structurally, each quantum theory should begin taking gravitational corrections at a different distance from the singularity. Because of its comparatively larger length scale, QED is more sensitive to the background geometry. It is conceivable that this may have an affect on the subsequent lengths scales, but is beyond the scope of this essay.

\section{Conclusions}
\renewcommand{\theequation}{5.\arabic{equation}} \setcounter{equation}{0}

Our current understanding says that quantum gravity effects become important when the curvature is of the order of Planck's scale \cite{Hawking-Ellis}. In the case of Schwarzschild black hole, the curvature is 
\begin{equation}
R_{\alpha \beta \gamma \delta } R^{\alpha \beta \gamma \delta } = \frac{48 \mu^2}{r^6}.
\end{equation}
Hence
\begin{equation}
\frac{48 \mu^2}{r^6} \approx \frac{1}{l_{Pl}^4}.
\end{equation}
or
\begin{equation}
r^3_{usual} \approx 
 2 \sqrt{3}\,\, r_S\,\, l_{Pl}^2
\end{equation}
Comparing it with our result for $l = l_{Pl}$, 
\begin{equation}
r^3_{our} >> r_S\,\, l_{Pl}^2. 
\end{equation}
This implies that our detailed calculation is in good agreement with the standard order-of-magnitude estimate. However, since we arrived at this conclusion by considering the tidal effect on the local inertial frame, our result suggests that within that characteristic radius, we cannot have a Lorentz frame and hence we cannot use the usual QFT in the Minkowski background. Also, this contradicts the SEP, i.e., the laws of physics will no longer reduce to that of SR. 
Perhaps, as is expected, the theory of quantum gravity may not have the singularities. In the case singularities are still present and is locally Lorentz invariant, then there will always be region around the singularity which cannot be explained by such a theory and hence will require a more fundamental theory to explain the physics inside of such a region. 
This begs the question as to if the correct theory of quantum gravity should be Lorentz invariant?

\section*{Acknowledgments}
We would like to thank Gerald Cleaver and Anzhong Wang for useful discussions.

\end{document}